# Colossal phonon drag enhanced thermopower in lightly doped diamond

Chunhua Li[1], Nakib H Protik[2], Pablo Ordejón[2] and David Broido[1*]

[1]Department of Physics, Boston College, USA

[2]Catalan Institute of Nanoscience and Nanotechnology, ICN2 (CSIC and BIST), Campus UAB,

08193 Bellaterra, Barcelona, Spain

**Abstract**-- Diamond is one of the most studied materials because of its unique combination of remarkable electrical, mechanical, thermal and optical properties. Using a fully self-consistent *ab initio* theory of coupled electron-phonon transport, we reveal another striking behavior: a huge drag enhancement of the thermopower of lightly doped diamond. Thermopower values of around 100,000 $\mu V\ K^{-1}$ are found at 100 K, significantly exceeding the highest previously measured value in the correlated metal $FeSb_2$, and occurring at much higher temperatures. The enormous thermopower in diamond arises primarily from exceptionally weak anharmonic phonon decay around and below 100 K that facilitates efficient momentum exchange between charge carriers and phonons through electron-phonon interactions. Exceedingly large thermoelectric power factors are also identified. This work gives insights into the physics of the coupled electron-phonon system in solids and advances our understanding of thermoelectric transport in the regime of strong drag.

# 1. Introduction

In metals and doped semiconductors, charge and heat currents are generated in response to driving temperature gradients and electric fields. The resulting electron and phonon transport are not independent because the charge carrier and phonon subsystems are coupled through electron-phonon interactions. This concept was first put forth by Peierls in 1930 [1] in response to Bloch's theory of electron transport [2], in which phonons were assumed to remain in equilibrium (Bloch assumption). In particular, the momentum transfer from the phonons to the charge carriers boosts the electrical current. This phenomenon, known as the phonon drag effect [3, 4], enhances the Seebeck thermopower, $S$, obtained experimentally as the ratio of the open-circuit electric field induced in response to the temperature gradient.

Phonon drag enhancements to $S$ were first observed in doped germanium [5, 6] and silicon [7] almost 70 years ago and have been found in many other materials since then. Metals typically have small $S$ and small phonon drag enhancements at low temperatures around and below 10K. Exceptions have been identified, most notably in iron antimonide (FeSb$_2$) for which a record $S$ value of 45 mV K$^{-1}$ was reportedly achieved at around 12 K [8]. Subsequent work attributed this large $S$ in FeSb$_2$ to phonon drag [9] while also finding more modest values of 16 mV K$^{-1}$ [10] and 27 mV K$^{-1}$ [11]. Large phonon drag enhanced thermopower values of around 25 mV K$^{-1}$ were also measured in isotopically enriched Ge at around 15K [12].

In this work, we report enormous phonon drag enhancement of the low temperature thermopower in lightly doped diamond, as predicted from first principles calculations. Thermopower values of around 100 mV K$^{-1}$ are found to be achievable near 100 K. These values are 100 times larger than those obtained when neglecting phonon drag. They are far higher than those reported previously in FeSb$_2$ [8-11] Ge [12], and also far higher than the measured [13] and calculated [14] values for diamond above 200 K. Correspondingly large values of the thermoelectric power factor, $PF = \sigma S^2$ ($\sigma$ is the electrical conductivity) are found that are much larger than the 2,300 $\mu$W K$^{-2}$ cm$^{-1}$ in FeSb$_2$, previously reported as a record high value [8], and orders of magnitude larger than those in the best thermoelectric materials such as Bi$_2$Te$_3$ and PbTe [15].

The extreme phonon drag enhancement of the low $T$ diamond thermopower is driven primarily by the exceedingly weak anharmonic decay of phonons, which promotes efficient transfer of quasi-momentum to the charge carriers even though electron-phonon interactions in diamond are relatively weak. The weak anharmonic phonon decay is also responsible for diamond's high phonon thermal conductivity, which we confirm precludes it from being a viable thermoelectric material in spite of its huge power factor, and even when considering phonon filtering concepts [16].

## 2. Theory

### 2.1. Predictive first principles approach to describe coupled carrier-phonon transport

Charge and heat conduction can be accurately described using the respective Boltzmann transport equations for electrons and phonons. However, most first principles transport calculations use the Bloch assumption i.e. they ignore the drag effects because of the computational expense needed to solve the coupled electron and phonon Boltzmann transport equations (eBTE/phBTE). Progress toward adjustable parameter free, *ab initio* implementations has been accomplished in recent years [14, 16-21], and phonon drag enhancements in measured thermopower data have been accurately described for Si [16-18, 21], GaAs [19] and diamond [14]. While extreme computational expense forced early *ab initio* approaches to adopt approximations, implementation of an efficient fully self-consistent first principles eBTE/phBTE solution to this long-standing and challenging transport problem has been recently developed [21]. The new computational code – elphbolt – allows predictive first principles calculations of thermopower even in the regime of strong phonon drag.

An applied electric field and temperature gradient create non-equilibrium electron and phonon distribution functions that cause charge and heat currents to flow. These distribution functions are calculated by solving the coupled eBTE/phBTE as implemented in elphbolt. From these distributions functions, the thermoelectric transport coefficients: electrical conductivity, $\sigma$, the electrical and phonon contributions to the thermal conductivity, $k_e$ and $k_{ph}$, the Seebeck thermopower, $S$, and the carrier and phonon contributions to the Peltier thermopower, $Q_c$ and $Q_{ph}$, are calculated. $Q_c$ and $Q_{ph}$ essentially measure the strengths of the induced heat currents of the charge carriers and phonons, respectively, due to an applied charge current. The Kelvin-Onsager relation [22, 23] connects the Seebeck and Peltier thermopowers as:

$$S = Q_c + Q_{ph} \qquad (1)$$

This fundamental relation of thermodynamics provides a powerful constraint that is rigorously satisfied in elphbolt.

The elphbolt code solves both the coupled and decoupled eBTE/phBTE for an applied electric field and temperature gradient. The difference between the calculated thermopowers for these two cases gives the magnitude of the drag enhancement. Solutions of these decoupled equations correspond to the generalized Bloch assumption where phonons (carriers) are taken to remain in equilibrium when considering carrier (phonon) transport. We then obtain the Seebeck thermopower from electron diffusion, $S_e$, i.e. the thermopower neglecting drag. In this approximation, $Q_{ph} = 0$, so the Kelvin-Onsager relation becomes: $S_e = Q_c$. The enhancement to the Seebeck thermopower due to phonon drag, $S_{drag}$, can then be defined as: $S_{drag} = S - S_e$.

Since $Q_c$ remains nearly unchanged in the presence of drag [16], the separability of the carrier and phonon contributions to the Peltier thermopower makes it straightforward to calculate the drag enhancement of the thermopower by solving the coupled eBTE/phBTE under applied electric field. As detailed in Ref. 21, this gives $Q_c = L_{21}/\sigma T$ and $Q_{ph} = K_{21}/\sigma T$ where

$$L_{21} = -\frac{2}{Vk_BT}\sum_{\nu} f_\nu^0(1-f_\nu^0)(\varepsilon_\nu - \mu)\mathbf{v}_\nu \otimes \mathbf{J}_\nu \tag{2}$$

$$K_{21} = -\frac{1}{Vk_BT}\sum_{\lambda} n_\lambda^0(n_\lambda^0+1)\hbar\omega_\lambda \mathbf{v}_\lambda \otimes \mathbf{G}_\lambda \tag{3}$$

$$\sigma = \frac{2e}{Vk_BT}\sum_{\nu} f_\nu^0(1-f_\nu^0)\mathbf{v}_\nu \otimes \mathbf{J}_\nu \tag{4}$$

In the equations, $\mathbf{J}_\nu$ and $\mathbf{G}_\lambda$ are the carrier and phonon deviation functions in response to the electric field, $f_\nu^0$ and $n_\lambda^0$ are the Fermi and Bose equilibrium distribution functions, $\nu$ designates the carrier band, $n$, and wavevector, $\mathbf{k}$, $\lambda$ designates the phonon branch, $j$, and wavevector, $\mathbf{q}$, $\varepsilon_\nu$ and $\hbar\omega_\lambda$ are the carrier and phonon energies, $\mathbf{v}_\nu$ and $\mathbf{v}_\lambda$ are the carrier and phonon velocities, $\mu$ is the chemical potential, $k_B$ is the Boltzmann constant, $T$ is the temperature, $e$ is the electron charge, and $V$ is the crystal volume.

## 3. Materials and Methods

First principles calculations were carried out within the framework of density functional theory using the local density approximation as implemented in the Quantum Espresso (QE) suite [24, 25]. We used the ultrasoft GBRV pseudopotentials [26] where the energy cutoff was set at 60 Ry. A lattice constant of 3.547 Å was used, which is within 1% of the measured value. Harmonic interatomic force constants (IFCs) were obtained within density functional perturbation theory in QE. The third-order anharmonic IFCs were computed using a 5×5×5 supercell using the finite displacement method [27]. We retained force constants up to the sixth nearest neighbors. To compute the electron-phonon scattering rates, we wannierized the four valence bands in diamond. Quantities in the real space (Wannier representation) were obtained from a coarse **k** grid of 12×12×12 and a **q** grid of 6×6×6 in EPW [28, 29], which were then transformed back to the reciprocal space (Bloch representation) on a fine **k** grid of 200×200×200 and a fine **q** grid of 100×100×100 for use in the the transport calculations within elphbolt.

## 4. Results

*4.1 Colossal thermopower in diamond*

Since doped diamond is *p*-type, we take holes as the carrier species and consider the dependence of the diamond thermopower as a function of the hole density, *p*, and *T*. Diamond has a cubic structure for which the thermoelectric transport coefficients are isotropic scalars. Figure 1 shows the calculated thermopower for natural diamond as a function of *T* for a hole density of $p = 10^{15}$ cm$^{-3}$ (red curve). The calculations include the following scattering mechanisms: (i) intrinsic hole-phonon scattering, (ii) phonon-hole scattering, (iii) anharmonic three-phonon scattering (iv) scattering of holes by charged impurities; (v) phonon-isotope scattering from the mass disorder produced by the 98.9% $^{12}$C and 1.1% $^{13}$C isotope mixture in natural diamond. Scattering of phonons by the mass variation of the dopant impurities is extremely weak for the density range considered and is neglected. Since scattering of phonons and holes by sample boundaries can significantly influence the transport at low temperatures, calculations in Fig. 1 are shown only down to 100K. The effect on the diamond thermopower from boundary scattering is considered below. The dotted black curve in Fig. 1 shows the calculated magnitude of the thermopower for lightly doped *n*-type silicon at a density of $2.8 \times 10^{14}$ cm$^{-3}$, obtained using the elphbolt code. This curve was shown to be in excellent agreement with the measured data of Ref. 21 at the same density.

Remarkably, the thermopower of natural diamond achieves a huge value approaching 100 mV K$^{-1}$, far larger than that measured in FeSb$_2$ (blue arrow) [8] and also far higher than the maximum measured value for isotopically enriched Ge (purple arrow) [12]. We note further that the huge diamond thermopower occurs at a temperature ten times higher than the peak values in FeSb$_2$ and Ge. As seen in Fig. 1, it also significantly exceeds thermopower for the Si sample, being over an order of magnitude larger at 100K. The effect of isotope enrichment will be discussed below.

Figure 2 shows the calculated $S$ as a function of $p$ for different values of $T$ between 100 K and 300 K, for natural diamond including all hole and phonon scattering mechanisms used for the red curve of Fig. 1. The 300K (green) curve shows good agreement with the measured value for low hole density (green circle, with error bars) from Ref. [13], as has also been found in prior *ab initio* calculations [14]. At 300 K, $S$ is enhanced by factors of between 2 to 3 compared with the case neglecting drag over the density range considered. Also, at 300 K, the measured $S$ is roughly twice that for Si at a similar density [7], consistent with our calculated results.

As $T$ is decreased the magnitude of the drag enhancement of $S$ increases. This increase is strikingly larger in diamond compared with that in Si. Specifically, while the calculated $S$ for Si shows a roughly three-fold increase in going from 300 K to 100 K, the corresponding increase for diamond is seen to be an order of magnitude larger. In comparison with the calculations at $T =$ 100 K in which phonons are constrained to remain in equilibrium (dashed red curve), $S$ is significantly enhanced when including phonon drag, ranging from a factor of about 20 at $p = 10^{18}$ cm$^{-3}$ to an astounding near one hundred-fold enhancement at $p = 10^{15}$ cm$^{-3}$. To our knowledge, this extreme drag enhancement of $S$ in diamond is far larger than that identified in any other material.

This enormous phonon drag enhancement of $S$ is primarily a consequence of diamond's strong covalent chemical bonding and light atoms. These features give a large phonon frequency scale (~40 THz), correspondingly large acoustic phonon velocities, and exceedingly weak anharmonic three-phonon scattering rates. Then, long-lived low frequency phonons are driven well out of equilibrium by the temperature gradient following which, through electron-phonon scattering, they can efficiently transfer momentum to the holes preferentially in the direction of the electric current.

In the Peltier picture for diamond at low $p$ and $T$, an applied electric field drives an electronic current under isothermal conditions. Electrons transfer momentum to low frequency acoustic phonons through electron-phonon scattering, thereby driving them out of equilibrium and establishing a phonon current in the same direction as the electron current. These phonons, being long-lived due to the weak phonon-phonon and phonon-electron scattering, give the dominant contributions to the thermopower. Thus, the thermal current generated by the electric field is primarily a phonon current.

Other factors can also be important in achieving large phonon drag enhancement of the thermopower. Specifically: (i) the electron-phonon coupling strength: a stronger electron-phonon coupling will enhance the transfer of momentum between carriers and phonons; (ii) not too high concentration of point defects: point defects obey the Rayleigh law that the phonon scattering rates are proportional to frequency to the fourth power. So, even for high concentrations of impurities, the scattering of low frequency phonons by point defects can still remain relatively weak; (iii) large enough grain sizes to reduce the boundary scattering of phonons: for small grain sizes, low frequency phonons are scattered reducing the drag enhancement of the thermopower, as discussed below and presented in Fig. 3.

The large drag effect occurs in spite of relatively weak electron-phonon interactions in diamond, as evidenced by the high measured hole mobility, with low density values of 2500 cm$^2$ V$^{-1}$ s$^{-1}$ (20,000 cm$^2$ V$^{-1}$ s$^{-1}$) at 300 K (100 K) [30]. The quantitative accuracy of this remarkable finding is supported by good agreement obtained between our calculated results and the measured diamond thermopower around and above room temperature [13] (See Fig. S1 in the Supplementary Information), which has also been achieved in recent *ab initio* calculations [14]. In addition, good agreement is obtained with measured mobility [30] and thermal conductivity [31-34] data for diamond across a wide temperature range (See Fig. S2 in the Supplementary Information).

To better understand this behavior, we note first that in semiconductors, low frequency phonons typically dominate the contributions to drag since they can satisfy the energy and momentum conservation in electron-phonon scattering processes, and three-phonon scattering rates tend to zero at low frequency minimizing dissipation of momentum from the electron-phonon system. The scattering rates of phonons by carriers increase roughly linearly with carrier density, and are only weakly dependent on $T$. At the same time, three-phonon scattering does not depend on $p$, and electron-phonon scattering is also relatively independent of carrier density (See Fig. S3 in Supplementary Information). Thus, for low enough carrier density, three-phonon scattering dominates over phonon-hole scattering, and $S_{drag}$ becomes independent of density [3, 16]. As $T$ decreases below 200 K, the three-phonon scattering rates plummet causing $S_{drag}$ to increase significantly so that $S \approx S_{drag} \gg S_e$, and thus $S$ becomes independent of $p$ as seen in Fig. 2 at low $p$ and low $T$. At high density, phonon-hole scattering dominates over the three-phonon scattering. Then, phonon scattering by carriers reduces the ability of the phonon current to transfer momentum to the carrier subsystem causing $S_{drag}$ to decrease with increasing carrier density. This behavior is known as the saturation effect [3, 16]. The evolving competition between phonon-phonon and phonon-carrier scattering as $p$ and $T$ change creates a transition in the behavior of $S_{drag}$, which is most evident in Fig. 2 at low $T$ where $S_{drag} \approx S$. The above description is further validated by examining the three-phonon scattering rates at 300 K and 100 K along with the phonon-hole scattering rates at $p = 10^{15}$ cm$^{-3}$ and $10^{18}$ cm$^{-3}$ for the low frequency phonons contributing to drag (Fig. S4 in Supplementary Information).

*4.2. Enhanced thermopower from isotope enrichment*

Prior *ab initio* calculations of the thermopower in diamond found almost no phonon drag enhancement upon isotopic enrichment of the carbon atoms [14]. Those calculated results extended only down to 200 K. The finding of large enhancement of the measured Ge thermopower around 15 K raises the question of whether such enhancement also occurs in diamond below 200K. To address this, we have performed thermopower calculations in which isotope enrichment is complete corresponding to a diamond crystal composed of 100 % $^{12}$C. Then, phonon-isotope scattering does not occur. The result is shown in blue dashed curve in Fig. 1. Above 200 K, three-phonon scattering rates are higher than phonon-isotope scattering rates, so isotopic enrichment produces negligible increase in $S$, as has been found previously [14]. However, below 200 K, the three-phonon scattering rates weaken so that they compete with the phonon-isotope scattering. Then, upon removal of the latter, $S$ is seen to increase, with an over a 50 % enhancement achieved by 100 K. This increase is a consequence of removing a scattering channel that dissipates momentum from the low frequency phonons that can now contribute to drag more efficiently. This enhancement is qualitatively consistent with the increase in $S$ found in isotopically enriched Ge [12].

*4.3. Effect of sample size on drag enhanced thermopower*

When the mean free paths of phonons and carriers contributing to drag become comparable to sample sizes, scattering from boundaries provides another source of dissipation. Figure 3 compares the $S(T)$ calculated for natural (1.1 % $^{13}$C) and isotopically enriched (0.1 % $^{13}$C) diamond with $p = 10^{15}$ cm$^{-3}$ for three different sample sizes: $L_b = 0.1$ mm, $L_b = 1$ mm and $L_b = 10$ mm. Here, the boundary scattering rates are approximated by phonon and carrier velocities divided by $L_b$: $\tau_b^{-1} = v/L_b$. If $L_b$ is large enough, then phonons are most influenced by the boundaries only at low $T$ where the exceedingly weak three-phonon scattering rates are comparable to the phonon-boundary scattering rates. We find that for $L_b \geq 1$ mm, values of $S$ above 100 K are relatively close to those ignoring the boundary scattering ($L_b = \infty$), while below 100 K, colossal values in the hundreds of mV K$^{-1}$ are found. Also, for $T$ below 100 K and $L_b = 10$ mm, isotope enrichment produces large increases in the thermopower. However, for small $L_b$ such as $L_b = 0.1$ mm, significant suppression of $S$ is seen across a wide range of temperatures, and isotope enrichment provides negligible benefit. Grain sizes smaller than 0.1 mm would give even stronger suppression extending to higher temperatures. These findings highlight the need for large grain sizes in order to achieve the extraordinarily high thermopower values in diamond.

*4.4. Huge thermoelectric power factor in diamond*

The extraordinarily large predicted $S$ for diamond invites consideration of the thermoelectric power factor, $PF = \sigma S^2$, which helps to characterize the efficiency of a material for thermoelectric cooling or power generation applications [15]. The best thermoelectric materials such as alloys of Bi$_2$Te$_3$ and PbTe typically have $PF$ values that are several tens of µW K$^{-2}$ cm$^{-1}$ [15]. The large measured $S$ in FeSb$_2$ around 10 K gave an enormous $PF$ of ~2300 µW K$^{-2}$ cm$^{-1}$, reported as a record high value [8]. We have found even larger $PF$ values for diamond at low $p$ and $T$. For example, at $p = 10^{16}$ cm$^{-3}$ and $T = 100$ K, a $PF$ of over 7,000 µW K$^{-2}$ cm$^{-1}$ is obtained for $L_b = \infty$, three times higher than the record value reported for FeSb$_2$ (Fig. S5 in Supplementary Information). Figure 4 shows that small $L_b$ suppresses the $PF$ across a wide range of temperatures, similar to the findings for the thermopower shown in Fig. 3. But, very large $PF$ is achieved for large $L_b$ at low $T$.

The weak phonon-phonon scattering responsible for the large $S$ and $PF$ in diamond is simultaneously responsible for the high $k_{ph}$ of diamond. $PF$ and $k_{ph}$ appear as a ratio in the thermoelectric figure of merit: $Z = PF/(k_e + k_{ph}) \approx PF/k_{ph}$. As a result, the calculated dimensionless figure of merit for diamond, $ZT$, remains at least a few hundred times smaller than the current largest measured values of $ZT \sim 1$ in the best thermoelectric materials. A route to achieving enhanced $ZT$ exploiting phonon drag [16] showed that a large fraction of the phonons contributing to $S_{drag}$ in silicon have frequencies below those contributing to $k_{ph}$. This offers the opportunity to boost $ZT$ by filtering out some phonons contributing to $k_{ph}$ (e.g. by selective phonon scattering through nanostructuring) while not affecting those lower frequency phonons contributing to $S_{drag}$. Fig. 5 shows the spectral frequency contributions to $S_{drag}$ ($Q_{ph}$) and $k_{ph}$ for diamond at 100 K and $p = 10^{15}$ cm$^{-3}$. Comparing the frequency ranges for the respective contributions shows that while perhaps 70 % of the phonons contributing to $k_{ph}$ could in principle be removed in the higher frequency range (i.e. > 4 THz) without suppressing $S_{drag}$, the necessary several hundred-fold enhancement of the diamond $ZT$ appears to be unfeasible using this approach. Isotopic enrichment produces a much larger enhancement in $k_{ph}$ than is true for $S_{drag}$, precluding any benefit to $ZT$ from this strategy.

## 5. Discussion

We note that the large drag enhancement predicted for the Peltier thermopower of diamond provides an increased ability to manipulate acoustic phonons with an electric field, a potentially desirable feature for phononics applications [35]. In addition, this prediction should be readily observable in experiments. Deep dopant levels (e.g. ~0.4 eV for boron) [36, 37] known to exist in diamond and a possible transition from band conduction to hopping transport at low $T$ [38] present synthesis and measurement challenges. However, alternative doping schemes provide possible routes to overcome these challenges [37, 39].

The colossal phonon drag enhancement of the thermopower found here for diamond raises the question of whether similar large enhancements might be found in other materials. Diamond is perhaps unique in its combination of stiff covalent bonding and light carbon atoms. These features contribute to the weak anharmonic phonon-phonon scattering, high thermal conductivity and large phonon drag enhanced thermopower in diamond. For materials that achieve high thermal conductivity in the same way as diamond, such as cubic boron nitride (BN) and boron phosphide (BP), we can expect large phonon drag enhanced thermopower. Future calculations and measurements will be required to assess the magnitude of such enhancements, but it is reasonable to speculate that values obtained for such materials will be smaller than those found here for diamond because the anharmonic scattering is weakest in diamond. In cubic boron arsenide (BAs) and similar materials, the high thermal conductivity is achieved by large contributions from high frequency acoustic phonons. However, as explained above, the largest phonon drag effects are achieved at low carrier density. In BAs, the transport-active carriers have small wave vectors and live close to the band edge. Then, the large phonon drag would come only from electron-phonon coupling between these carriers and low frequency acoustic phonons. The low frequency phonons in BAs have similar anharmonic scattering rates to those in Silicon. Therefore, we should expect the phonon drag effects in BAs to be similar to those in Silicon and much lower than those in diamond.

## 6. Conclusions

Using a fully self-consistent theory of coupled carrier-phonon transport, a colossal thermopower is identified in lightly-doped diamond. The large thermopower results from extraordinary drag enhancement. Thermopower values of around 100 mV K$^{-1}$ are found at 100 K, which are far larger than those measured in other materials and also occur at remarkably higher temperatures. The behavior stems primarily from a collapse in anharmonic phonon-phonon scattering rates that dominates the low temperature thermoelectric transport behavior of the carrier-phonon system for low carrier densities.

These findings reveal another remarkable property of diamond, along with its ultrahigh hardness and thermal conductivity and promise as an opto-electronic material, and they give fundamental insights into the nature of coupled carrier-phonon transport in the regime of extreme drag.

**Acknowledgements**

This work was supported primarily by the US Department of Energy (DOE), Office of Science, Basic Energy Sciences (BES) under award # DE-SC0021071 (first principles calculations of diamond electron and phonon scattering rates, thermopower, electrical and thermal conductivities, and power factor). Work at ICN2 was supported by EU-H2020 through H2020-NMBP-TO-IND project GA n. 814487 (INTERSECT), the Severo Ochoa program from Spanish MINECO (Grant No. SEV-2017-0706), Spanish MICIU, AEI and EU FEDER (Grant No. PGC2018-096955-B-C43) and the CERCA Program of Generalitat de Catalunya. D.B. and C.L. acknowledge the Boston College Linux clusters for computational resources and support. D.B. acknowledges helpful discussions with Prof Li Shi from the University of Texas, Austin and Prof. Zhifeng Ren from the University of Houston.

**Author Contributions**

D.B. and C.L. designed project; C.L. performed the calculations; D.B. and C. L. analyzed data with support from N.H.P. and P.O.; N.H.P. provided software support; D.B. wrote the paper; all authors contributed to manuscript revisions.

**Figures**

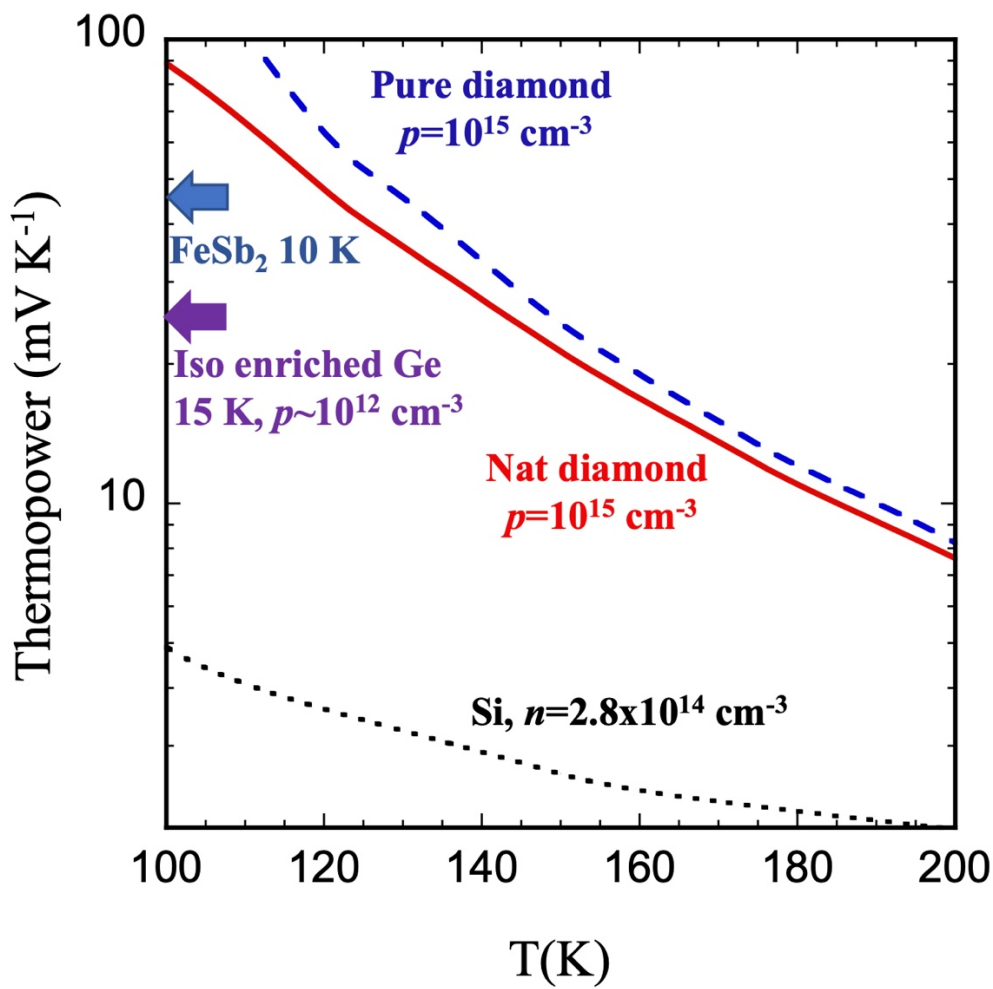

**Figure 1**: Temperature dependence of thermopower of diamond with hole density $p = 10^{15}$ cm$^{-3}$, calculated from first principles. Solid red curve is for natural diamond (1.1% $^{13}$C). Dashed blue curve is for isotopically pure diamond (100 % $^{12}$C). Dotted black curve is calculated magnitude of the thermopower for silicon at electron density $2.8 \times 10^{14}$ cm$^{-3}$ [21]. Blue arrow shows the maximum measured thermopower for FeSb$_2$ at 10 K [8], while purple arrow shows the maximum measured thermopower for isotopically enriched Ge at 15 K [12].

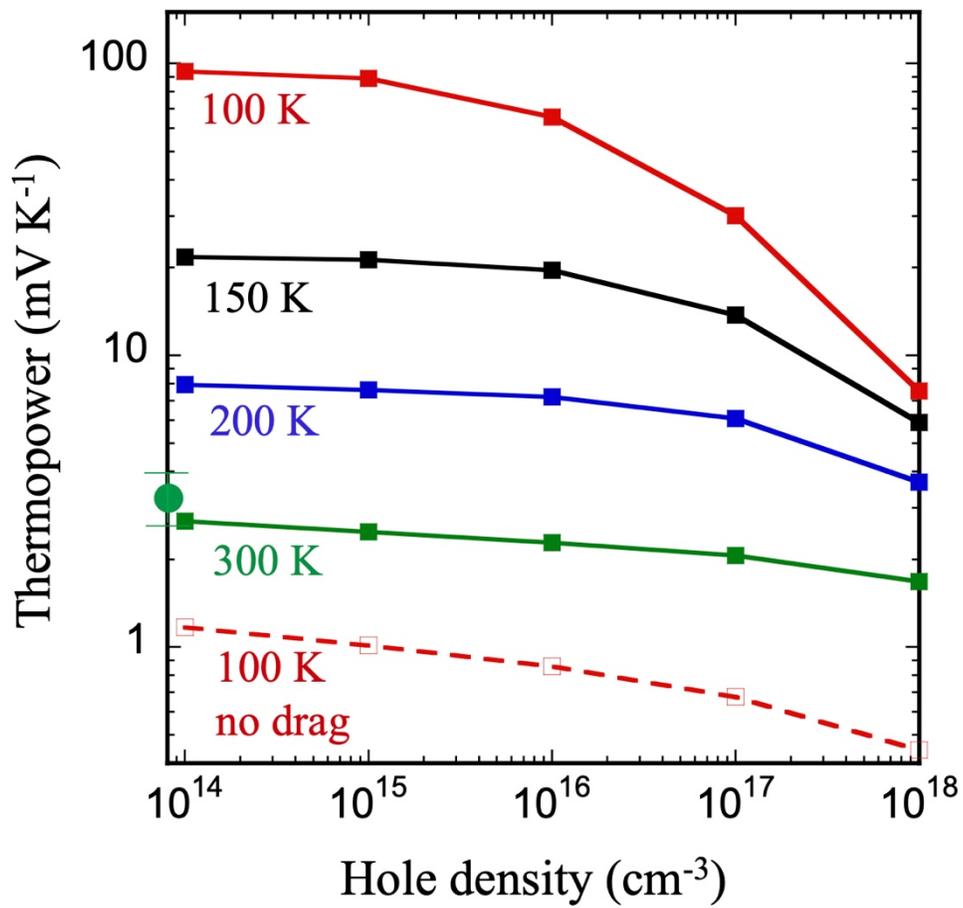

**Figure 2**: Calculated thermopower for natural diamond as a function of hole density for different temperatures. Solid curves include phonon drag, while dashed red curve neglects it. Green point is measured data at 300 K from Ref. 13.

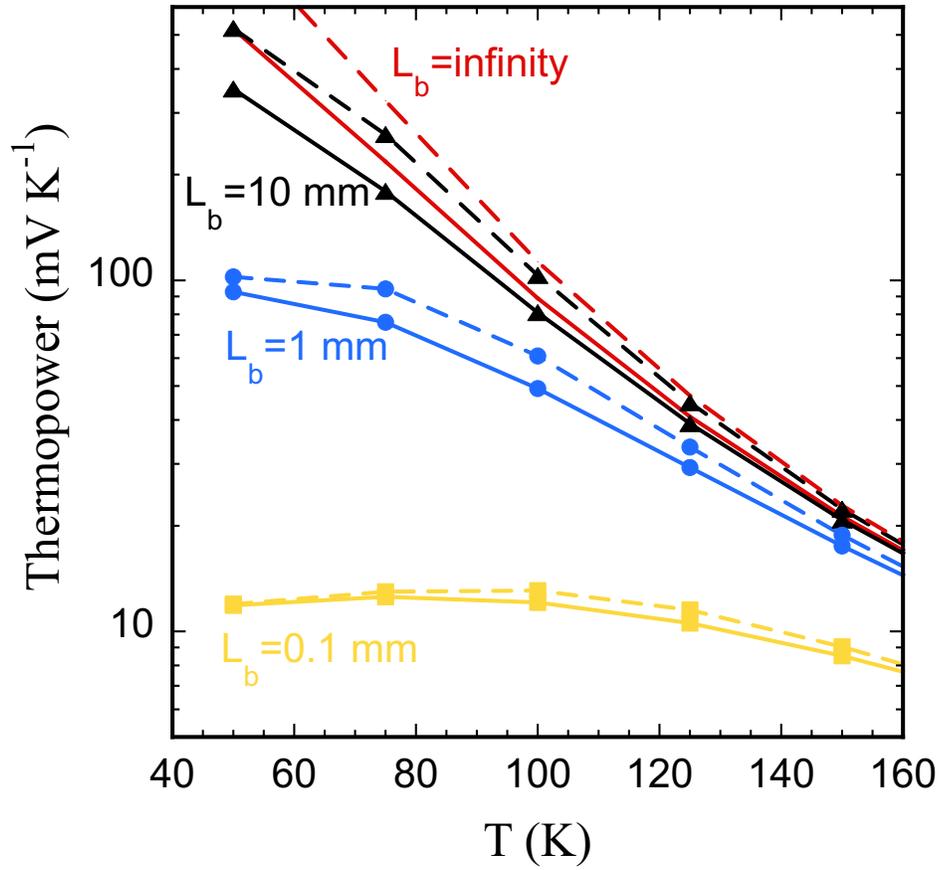

**Figure 3**: *Ab initio* calculations of diamond thermopower for hole density $10^{15}$ cm$^{-3}$ as a function of temperature for different crystal grain sizes, $L_b$. Solid and dashed curves are for natural diamond (1.1 % $^{13}$C) and isotopically enriched diamond (0.1 % $^{13}$C), respectively.

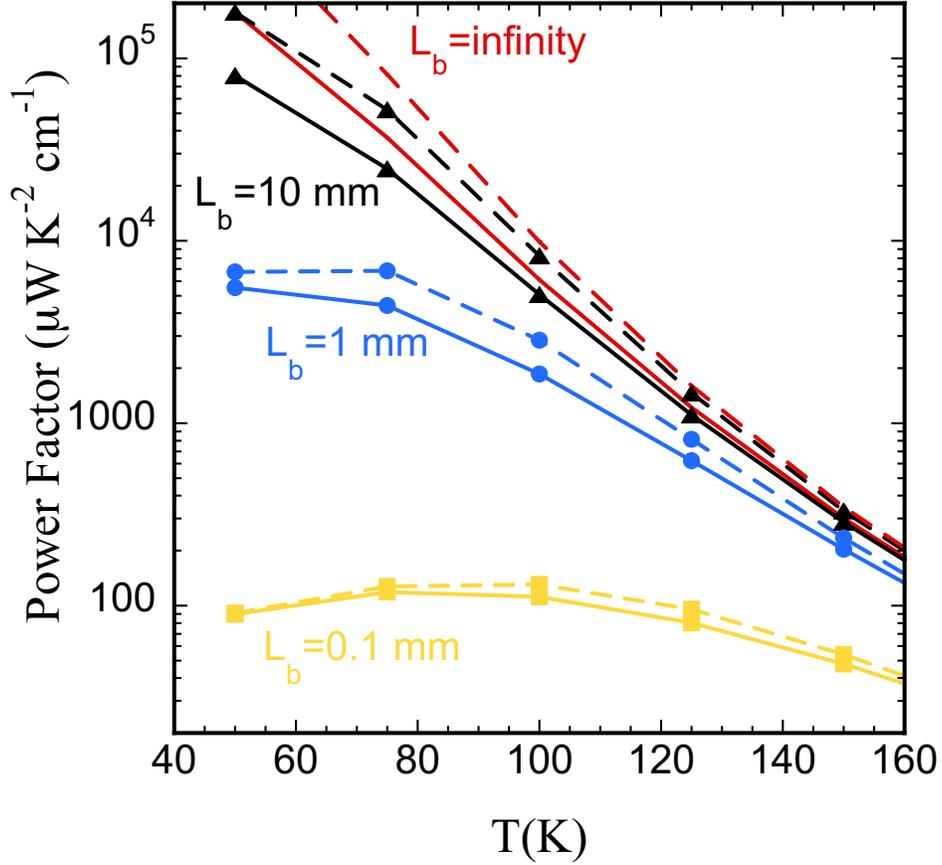

**Figure 4:** Thermoelectric power factor for diamond for $p = 10^{15}$ cm$^{-3}$, including boundary scattering with $L_b$ = 0.1 mm, 1 mm, 10 mm and $L_b = \infty$. Solid curves are for natural diamond with 1.1% $^{13}$C. Dashed curves are for isotopically enriched diamond with 0.1% $^{13}$C.

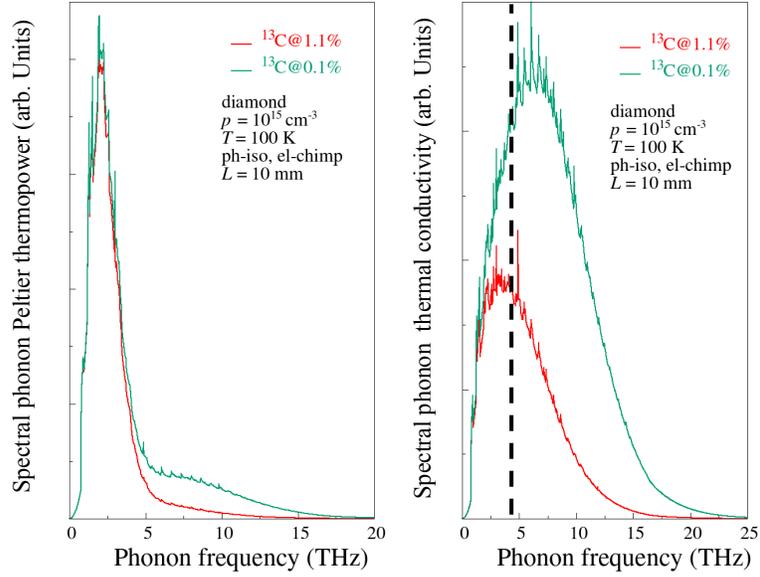

**Figure 5:** (a) Spectral contributions to drag thermopower, $Q_{ph}$, for natural (red) and isotopically enriched (green) diamond with $p = 10^{15}$ cm$^{-3}$ at a temperature of 100 K and for a boundary scattering size of 10 mm. (b) Spectral contribution to the phonon thermal conductivity, $k_{ph}$. Dashed vertical black line shows frequency below which phonons contribute to $Q_{ph}$.

**Supplementary Information for**

Colossal phonon drag enhanced thermopower in lightly doped diamond

Chunhua Li[1], Nakib H Protik[2], Pablo Ordejón[2] and David Broido[1]

[1]Department of Physics, Boston College, USA

[2]Catalan Institute of Nanoscience and Nanotechnology, ICN2 (CSIC and BIST), Campus UAB, 08193 Bellaterra, Barcelona, Spain

Supplementary data, containing plots of calculated diamond Seebeck thermopower, mobility and thermal conductivity compared to measured data, electron and phonon scattering rates, and thermoelectric power factor.

**Comparison of calculated diamond thermopower to measured data**

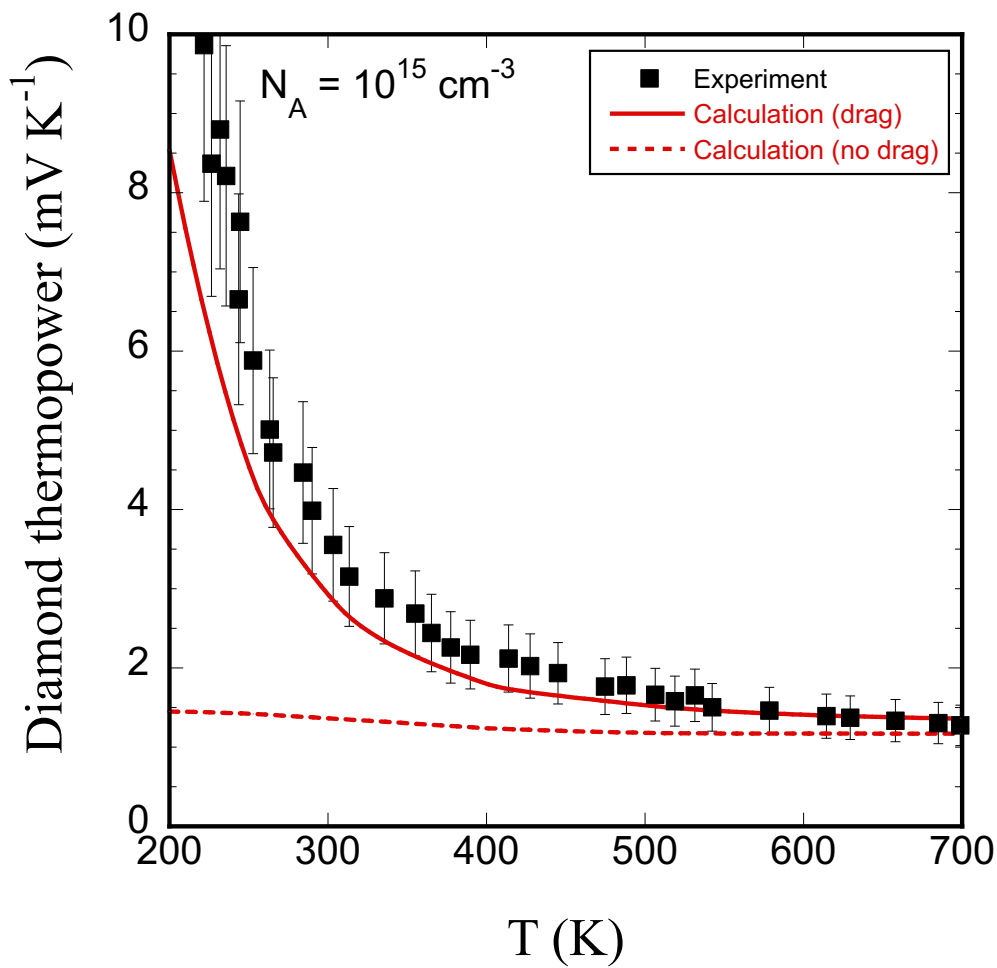

**Figure S1:** Calculated Seebeck thermopower for boron-doped diamond including drag effects (solid red curve) and omitting them (dashed red curve) compared to experimental thermopower measurements from Goldsmid [1] (black squares with error bars). For the calculations, the density of boron acceptor dopant atoms is taken to be $N_A = 10^{15}$ cm$^{-3}$. The hole concentration depends on temperature and acceptor ionization energy. This is accounted for as in the calculations of Ref. 2.

**Comparison of calculations to measured mobility and phonon thermal conductivity data**

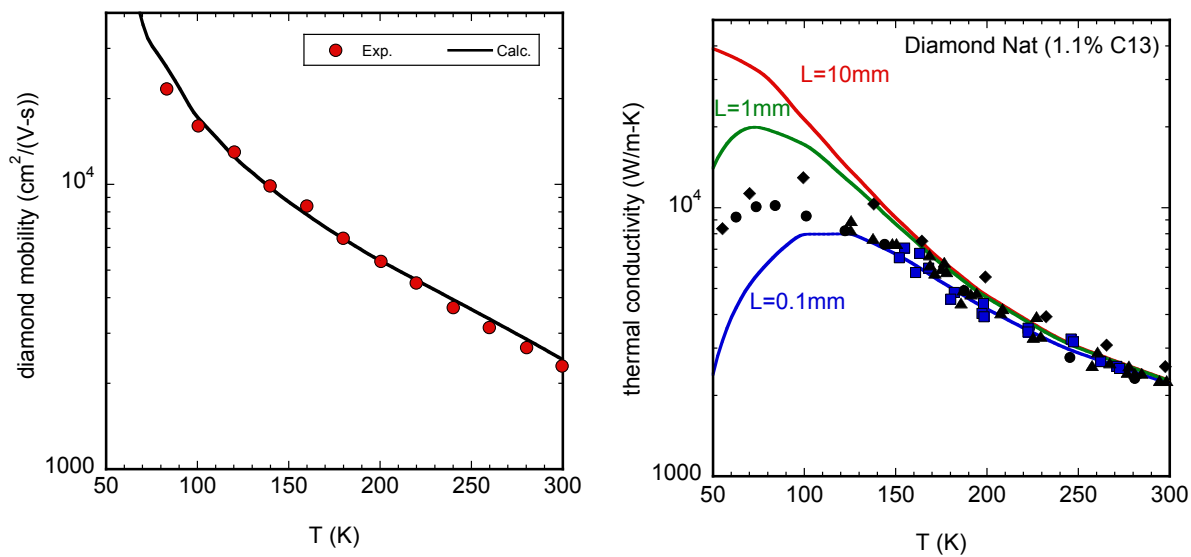

**Figure S2:** (a) Measured low hole density mobility in diamond .vs. temperature [3] compared with ab initio calculated values. A hole density of $10^{15}$ cm$^{-3}$ has been taken in the calculations, and scattering from charged impurities has not been included consistent with the measurement technique. (b) Measured temperature dependent thermal conductivity data for natural diamond [4-7], compared with first principles calculations. Calculations include three-phonon scattering, phonon-isotope scattering with 1.1 % $^{13}$C and scattering of phonons from boundaries, with three different sample sizes considered: L = 0.1 mm, L = 1 mm and L = 10 mm. Good agreement with the measurements is achieved above 150 K, where the effect of sample dependent boundary scattering is reduced.

**Hole-phonon scattering rates for diamond at two different temperatures and densities**

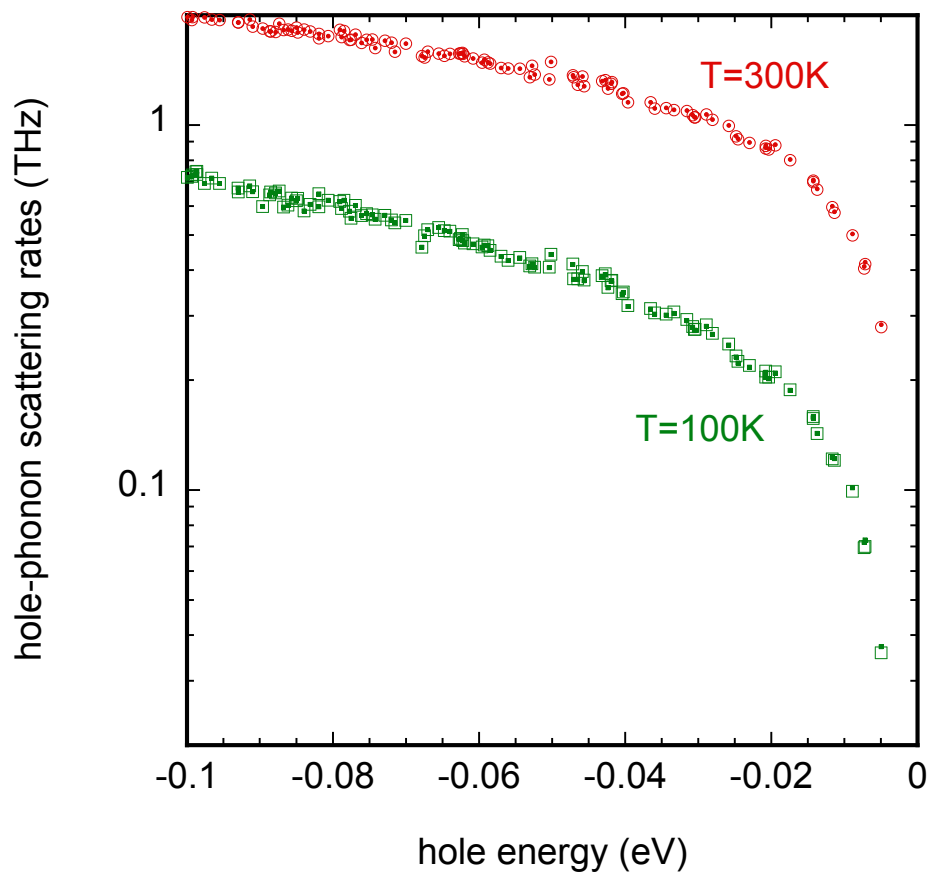

**Figure S3:** Hole-phonon scattering rates for diamond from first principles calculations for 300 K (red circles and dots) and 100 K (green squares and dots), and for densities of $p = 10^{15}$ cm$^{-3}$ (open red circles and green squares) and $p = 10^{18}$ cm$^{-3}$ (red and green dots). The zero of energy is set at the valence band maximum. The scattering rates depend strongly on $T$ but are roughly independent of $p$.

**Phonon scattering rates in natural diamond at different temperatures and hole densities**

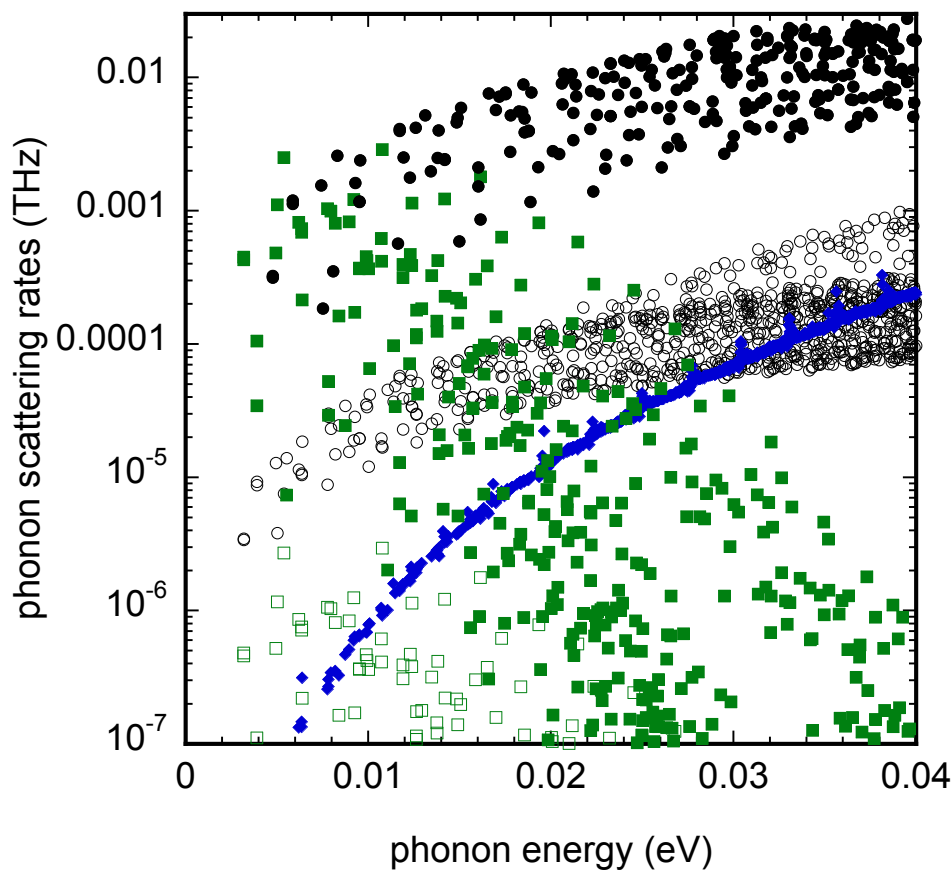

**Figure S4:** Intrinsic phonon scattering rates for diamond in the region of low frequency where phonons contribute to drag. The solid and open black circles show the three-phonon scattering rates at 300 K and 100 K, respectively. The solid and open green squares show the phonon-hole scattering rates at 100 K for $p = 10^{18}$ cm$^{-3}$ and $p = 10^{15}$ cm$^{-3}$, respectively. Starting from $T = 300$ K and $p = 10^{18}$ cm$^{-3}$, where both the three-phonon scattering rates and phonon-hole scattering rates are relatively strong, lowering $T$ from 300 K to 100 K reduces the three-phonon scattering rates by two orders of magnitude. At $T = 100$ K and in the range between $p = 10^{18}$ cm$^{-3}$ and $p = 10^{16}$ cm$^{-3}$, scattering of phonons by holes dominates, and then S increases substantially with decreasing hole density. Below $p = 10^{16}$ cm$^{-3}$, the phonon-phonon scattering again dominates the transport, giving the huge phonon drag enhanced thermopower (Figs. 1 and 2, main text), and $S$ saturates, becoming independent of $p$ (red curve in Fig. 2, main text). Blue points show the phonon-isotope scattering rates, for natural diamond (1.1% $^{13}$C), which are independent of $T$.

**Thermoelectric power factor for diamond for different temperatures and hole densities**

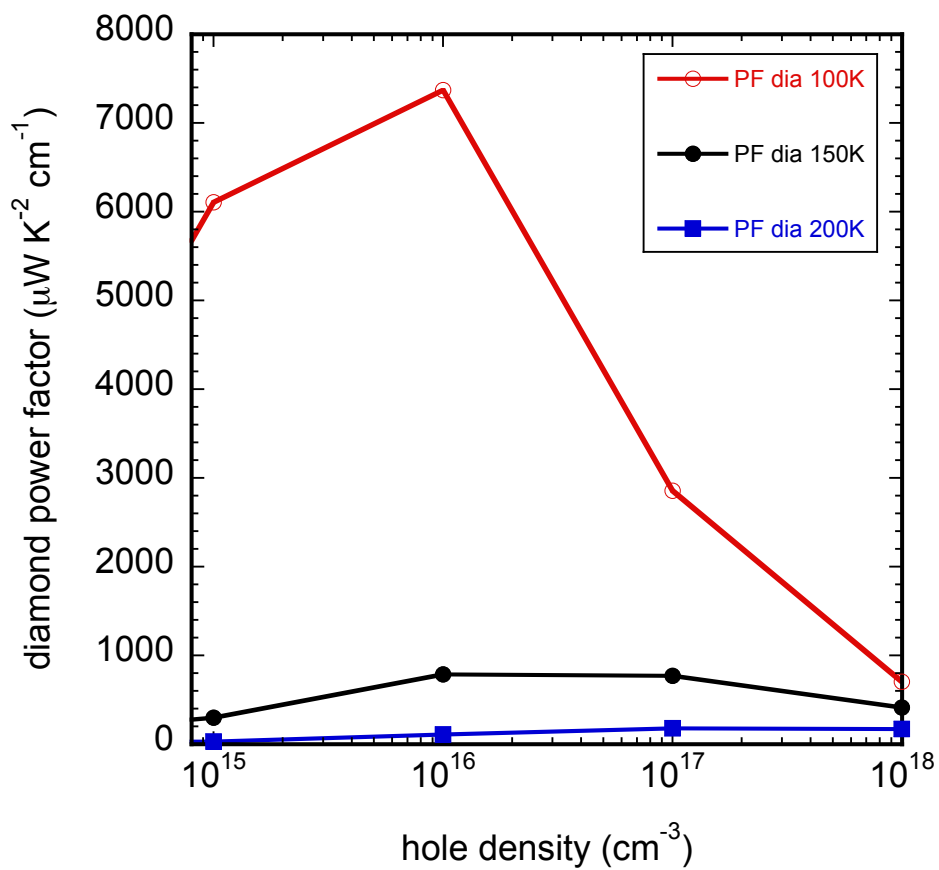

**Figure S5:** Calculated thermoelectric power factor for natural diamond as a function of hole density for different temperatures. Calculations include electron-phonon, phonon-electron, three-phonon, phonon-isotope and carrier-ionized impurity scattering. Boundary scattering has not been included.